\documentclass[aps,superscriptaddress,preprint,nofootinbib]{revtex4}

\usepackage{amsmath}
\usepackage{graphicx}
\usepackage{amssymb}
\usepackage{dcolumn}
\usepackage{bm}
\usepackage{color}

\makeatletter

\begin{document}

\title{Hydrodynamics of chiral squirmers}

\author{P. S. Burada} \thanks{\tt Corresponding author:psburada@phy.iitkgp.ac.in}
\affiliation{Department of Physics, Indian Institute of Technology Kharagpur, Kharagpur 721302, India}

\author{R. Maity}
\affiliation{Department of Physics, Indian Institute of Technology Kharagpur, Kharagpur 721302, India}

\author{F. J\"ulicher}
\affiliation{Max Planck Institute for the Physics of Complex Systems,  
N\"othnitzer Str. 38, 01187 Dresden, Germany}

\date{\today}

\begin{abstract}
Many microorganisms take a chiral path while swimming in an ambient fluid. In this paper, we study the combined behavior of two chiral swimmers using the well-known squirmer model taking into account chiral asymmetries. In contrast to the simple squirmer model, which has an axisymmetric distribution of slip velocity, the chiral squirmer has additional asymmetries in the surface slip, which contribute to both translations and rotations of the motion. As a result, swimming trajectories can become helical and chiral asymmetries arise in the flow patterns. We study the swimming trajectories of a pair of chiral squirmers that interact hydrodynamically. This interaction can lead to attraction and repulsion, and in some cases even to bounded states where the swimmers continue to periodically orbit around a common average trajectory. Such bound states are a signature of the chiral nature of the swimmers. Our study could be relevant to the collective movements of ciliated microorganisms.
\end{abstract}

\maketitle

\section{Introduction}

Understanding the locomotion and the swimming behavior of biological microswimmers has drawn substantial attention of scientists \cite{Lighthill, Blake, brennen, yates, berg, laugap, Eric_book}. There exists a rich literature exploring the novel phenomena emerging from cellular motility at the sub-millimeter scale \cite{pinel, holwill, jahn}; nutrient transport \cite{magar, michelin}, stochastic dancing of \textit{Volvox} algae \cite{drescher}, formation of biofilm \cite{pratt, houry, berke, li}, stochastic swimming \cite{berg1, howse, romanczuk}, swimming of sperm cells \cite{berg, FriedrichPNAS2007, FriedrichPRL2009} are a few examples. 
These studies have inspired the design of artificial microswimmers for microsurgery and targeted drug delivery \cite{nelson}. 
Different microswimmers employ different motility mechanisms to swim in a fluid. 
We are motivated by motion based on ciliary propulsion of many microorganisms.

Many ciliated microorganisms can actively swim due to the periodic motion of hair-like appendages such as motile cilia attached on their surface \cite{Bray}. Swimming corresponds to a net motion, relative to the background fluid, resulting from non-reciprocal shape changes of the swimmer \cite{Taylor,Purcell,Wilczek}. The beating of cilia on the surface of organisms such as \textit{Paramecium} or \textit{Opalina} generates intricate wave patterns, called metachronal waves, which enable the body to swim in a fluid \cite{Machemer}. In addition to biological swimmers, artificial microswimmers have recently attracted much interest. Several artificial microswimmers use different self-propulsion mechanisms that involve the generation of surface flows \cite{howse,Leonetti, Lammert,Golestanian1,Vilfan, liu, saadat}.

The hydrodynamics of the resulting propulsion was studied with a simple model called squirmer, 
introduced by Lighthill \cite{Lighthill} and further developed by Blake and others \cite{Blake, Pedley}. A squirmer is a spherical object moving in a fluid, driven by a pattern of slip velocity on its surface.
Such a squirmer is force free, but is associated with a force dipole. 
As a result, the squirmer can swim in the fluid and generate a characteristic hydrodynamic flow pattern. 
Though the squirmer was first developed as a model to understand ciliary propulsion, it 
is now widely used to study other types of microswimmers broadly classified as pullers and pushers \cite{laugap,wu}. 
While pullers have an extensile force dipole, resulting, e.g., from the front part of the body, pushers have a contractile force dipole stemming, e.g., from the rear part of the body \cite{Eric_book}.

In simple versions of the squirmer model, the surface slip of the squirmer is axisymmetric \cite{Lighthill, Blake} which leads to swimming in a direction along the axis of symmetry. 
Many microswimmers, for example, \textit{Marine Zooplankton}, propel along a chirally asymmetric path and generate flow field non-axisymmetric in nature. Such chiral swimmers exhibit rotational motion and often move along helical paths \cite{Gray95,Brokaw58,Crenshaw96,FriedrichPNAS2007,FriedrichPRL2009,Jekely}. A recent experimental study shows that artificially designed swimmers which are generating chiral flow, propel along a helical path in a surfactant solution \cite{yamamoto}. Understanding the role of chirality in the motility of microswimmers has therefore drawn significant attention \cite{yamamotosoft, yamamotopre, fadda, tjhung, carenza, lancia, pak}. Many biological phenomena, e.g., plankton bloom in ocean \cite{hallegraeff}, nutrient uptake by swimming organisms, bioconvection \cite{pedleybio, pedleybio2}, transient clustering \cite{llopis}, cancer and tissue development \cite{friedl}, amoebae aggregation due to starvation \cite{tyson} result from the collective motion of microorganisms and motile cells. 
For example, hydrodynamic bound states have been discussed when a pair of bottom-heavy \textit{Volvox} swimming near a substrate \cite{drescher}.
Therefore, it is important to understand the results of interactions between swimmers for example via the generalized flows. 

In this paper, we briefly discuss the general chiral squirmer model and derive the solution of the hydrodynamic flow field around the body. 
We then study the motion of a pair of chiral squirmers that exhibit complex swimming trajectories as they interact via the hydrodynamic flow field.
The rest of this article is organized as follows.  In section~\ref{sec:model} we introduce the chiral squirmer model and the governing equations. Hydrodynamics of a single chiral squirmer is discussed in section~\ref{sec:1chsquirmer}. The combined behavior of two chiral squirmers is presented in section~\ref{sec:2chsquirmer}. The main conclusions are provided in section~\ref{sec:conclusions}.

\section{Hydrodynamic flow of a squirmer}
\label{sec:model}

To discuss the swimming of microorganisms in the low Reynolds number regime, inertial forces can be neglected. In an incompressible, Newtonian fluid, 
the hydrodynamic flow field obeys the Stokes equation \cite{Happel}, 
\begin{align}
\label{eq:stokes}
\eta \nabla^2 {\bf u} = {\bf \nabla}p \, , 
\end{align}
where $\eta$ is the viscosity, ${\bf u}$ is the velocity field, 
and $p$ is the pressure field which plays the role of a Lagrange multiplier to impose the 
incompressibility constraint $ {\bf \nabla} \cdot {\bf u} = 0$.
\begin{figure}[htb]
  \centering
  \includegraphics[scale=0.6]{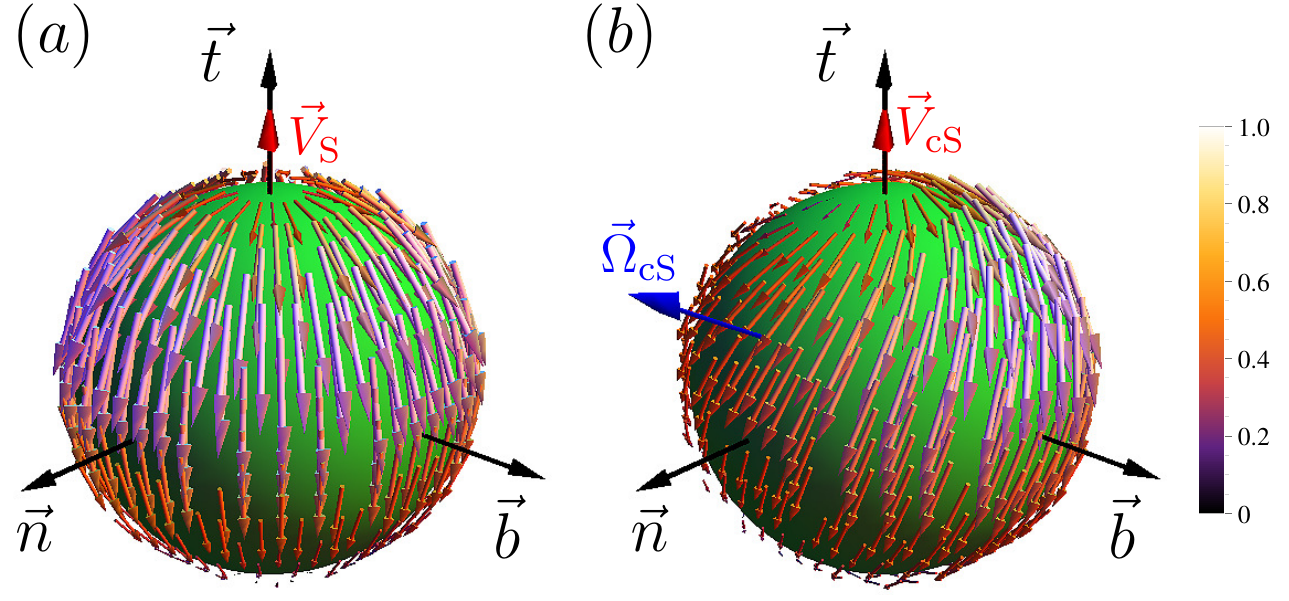}
  \caption{(Color online) 
    Examples of surface slip velocity patterns  
    of a simple squirmer (a) and a chiral squirmer (b) 
    in the body-fixed reference frame {\bf n}, {\bf b}, and ${\bf t}$.
    Parameter values are
    $ {\bf V} = v(0,0,1)$, 
    $ {\bf \Omega} = v(1/\sqrt{2}, 0, 1/\sqrt{2})/a$, and
    $\beta_{2 0}^r = v/3$. 
    For the simple squirmer (axisymmetric), $\gamma_{2 0}^r = 0$ and for the chiral squirmer $\gamma_{2 0}^r = v/3$. All other coefficients are set to zero.}
  \label{fig:slip}
\end{figure}

A squirmer is a rigid spherical body of radius a. 
On its surface, we prescribe a surface slip velocity ${\bf S} (\theta,\phi)$ which is tangential to the surface and parameterized by the polar and azimuthal angle $\theta$ and $\phi$, respectively, in a body-fixed frame. The latter is defined by three orthogonal unit vectors attached to the sphere center ${\bf n}$, ${\bf b}$, and ${\bf t}$ see Fig.~\ref{fig:slip}. 
It is convenient to express this surface slip pattern using gradients of spherical harmonics that form a basis for tangential vectors on the surface \cite{Happel}. 
The slip velocity can then be expressed in the form 
\begin{align}
\label{eq:slip}
{\bf S}(\theta, \phi) = \sum_{l=1}^{\infty}\sum_{m= -l}^l
\Big[-\beta_{lm}\, {\boldsymbol \nabla}_s \left( P_l^m(\cos\theta) \,e^{i m \phi} \right) + \gamma_{lm}\, {\bf \hat{r}} \times {\boldsymbol \nabla}_s \left( P_l^m(\cos\theta) \,e^{i m \phi} \right) \Big]\,, 
\end{align}
where ${\boldsymbol \nabla}_s$ is the gradient operator on the surface of the sphere defined as 
${\boldsymbol \nabla}_s = {\bf e}_\theta \, {\partial}/{\partial}\theta +(1/\sin\theta)\, {\bf e}_\phi {\partial}/{\partial}\phi$, 
${\bf \hat{r}}$ is the unit vector in radial direction, 
$P_l^m(\cos\theta) \,e^{i m \phi}$
are non-normalized spherical harmonics, where $P_l^m(\cos\theta)$ denotes 
Legendre polynomials.
The complex coefficients $\beta_{l m}$ and $\gamma_{l m}$ are the mode amplitudes of the prescribed surface slip velocity. 
We introduce the real and imaginary parts of these amplitudes 
as $\beta_{l m} = \beta_{l m}^r + i\,m\, \beta_{l m}^i$ and $\gamma_{l m} = \gamma_{l m}^r + i\,m\, \gamma_{l m}^i$ 
with complex conjugates $\beta_{l m}^\ast = (-1)^m \beta_{l, -m}$ and $\gamma_{l m}^\ast = (-1)^m \gamma_{l, -m}$, respectively.

The velocity and rotation rate can be calculated directly from the surface slip profile Eq.~(\ref{eq:slip}) \cite{Stone}. 
They can be expressed in the body fixed reference frame as 
${\bf V} = 2(\beta_{11}^r , \, \beta_{11}^i , \, \beta_{10}^r )/3$ and
${\boldsymbol \Omega} = (\gamma_{11}^r, \, \gamma_{11}^i, \, \gamma_{10}^r )/a$,
respectively.
Without loss of generality, we can choose the body-fixed reference frame $({\bf n}, {\bf b}, {\bf t})$ 
such that ${\bf t}$ points in the direction of motion.
With this choice, we have $\beta_{11}^r = \beta_{11}^i = 0$ 
and we write $\beta_{1 0}^r = 3 v/2$ such that $v=|{\bf V}|$ is the speed of the swimmer. The translation velocity, the rotation rate, and the rate of energy dissipation $Q$ in the flow field can then be expressed as \cite{Stone},
\begin{align}
  \label{eq:vel_trans}
  {\bf V} & = v \,  {\bf t} \ , \\ 
  \label{eq:vel_rot}
  {\bf \Omega} & =  \frac{\gamma_{11}^r}{a} \,{\bf n} +  \frac{\gamma_{11}^i}{a} \,{\bf b} +  \frac{\gamma_{10}^r}{a} \, {\bf t} \ , \\
  \label{eq:dissipation}    
 Q & = 12 \pi  a \eta \,\left(v^2+ \frac{4 a^2}{9} |{\bf \Omega}|^2  \right)  \,.
\end{align}
Eq.~(\ref{eq:dissipation}) reveals that a chiral squirmer dissipates more energy 
than a non-chiral one due to additional flows associated with rotations. 

In Eq.~(\ref{eq:slip}), the modes of surface slip described by $\gamma_{lm}$ (for $m \neq 0$) break the axial symmetry of the flow profile. 
For the case $\gamma_{lm} = 0$, the swimmer is non-chiral because modes corresponding to the coefficients $\beta_{lm}$ do not generate an azimuthal part of the flow field. Considering only the modes $\beta_{l 0}$  in Eq.~(\ref{eq:slip}), setting all other modes to zero, one recovers the axisymmetric squirmer model which generates translational motion only \cite{Lighthill, Blake}. Fig.~\ref{fig:slip}(a) shows an example of such an axisymmetric surface slip pattern. 
The former pattern is axisymmetric concerning the $\bf t$ axis. 
An example of the asymmetric surface slip of a chiral squirmer 
is shown in Fig.~\ref{fig:slip}(b).

\section{Hydrodynamics of a single chiral squirmer} 
\label{sec:1chsquirmer}

\subsection{Hydrodynamic flow field} 
\label{hydrodynamic field}

For the prescribed surface slip of the chiral squirmer, 
we can calculate the corresponding flow field. 
Using a lab reference frame (lf) which is at rest with respect 
to the fluid away from the swimmer, 
the flow field, the pressure field, and the vorticity are obtained as
\begin{align}
\label{eq:VelocityField}
{\bf u}_\mathrm{lf}({\bf r})  
   & = \frac{3 v}{2} \frac{a^3}{r^3} 
   \left[ P_1({\bf t} \cdot {\bf \hat{r}})\,{\bf \hat{r}} - \frac{\bf t}{3} 
   \right]   
+ 3\,\beta_{2 0}^r \left[ \frac{a^4}{r^4} - \frac{a^2}{r^2} \right]\,
P_2({\bf t} \cdot {\bf \hat{r}})\, {\bf \hat{r}}  \nonumber \\
& + \beta_{2 0}^r \frac{a^4}{r^4} 
P_2^{\prime}\left( {\bf t} \cdot {\bf \hat{r}} \right) [ ({\bf t} \cdot {\bf \hat{r}}) {\bf \hat{r}} - {\bf t} ] 
- \gamma_{2 0}^r\,\frac{a^3}{r^3} \, 
P_2^{\prime}\left( {\bf t} \cdot {\bf \hat{r}} \right)
{\bf t} \times {\bf \hat{r}} \,,\\ 
\label{eq:pressure_LB}
  p_\mathrm{lf}({\bf r})
  & =  - 2 \eta\,\beta_{2 0}^r \,\frac{a^2}{r^3} \,P_2\left({\bf t} \cdot {\bf \hat{r}}\right) \,,\\
 \label{eq:VorticityField}
\boldsymbol{\omega}_\mathrm{lf}({\bf r})  &= \frac{{\boldsymbol \nabla} \times {\bf u}_\mathrm{lf}({\bf r})}{2}\,,
\end{align}
where 
${\bf t}$ is the swimming direction, 
$r$ is the distance from the center of the swimmer where the flow field is determined, 
${\bf \hat{r}} = {\bf r}/r$ is the radial vector,
$P_2(x)$ denotes a second-order Legendre polynomial, and 
$P_2^{\prime} = dP_2/dx$ with $x = \mathbf{t}\cdot \mathbf{\hat{r}} = \cos \theta$.
Note that in Eq.~(\ref{eq:VelocityField}) we have only written terms up to $l = 2$ and do not consider higher order contributions, since they decay more rapidly with $r$. 
Besides, to have a minimal model, we have ignored $l = 2$ modes with $m\neq 0$. 
However, it is straightforward to include the additional terms in the analysis. 
Note that in Eq.~\ref{eq:VelocityField}, the terms $o(1/r^4)$ decay faster than lower order terms, and hence their contribution is negligible to the flow field at a large distance.
The flow field in the body frame (bf) can be obtained from that in the lab frame (lf) as 
${\bf u}_\mathrm{bf}({\bf r}) = {\bf u}_\mathrm{lf}({\bf r}) - {\bf V} - {\bf \Omega} \times {\bf r}$. 

\begin{figure}[t!]
  \centering
  \includegraphics{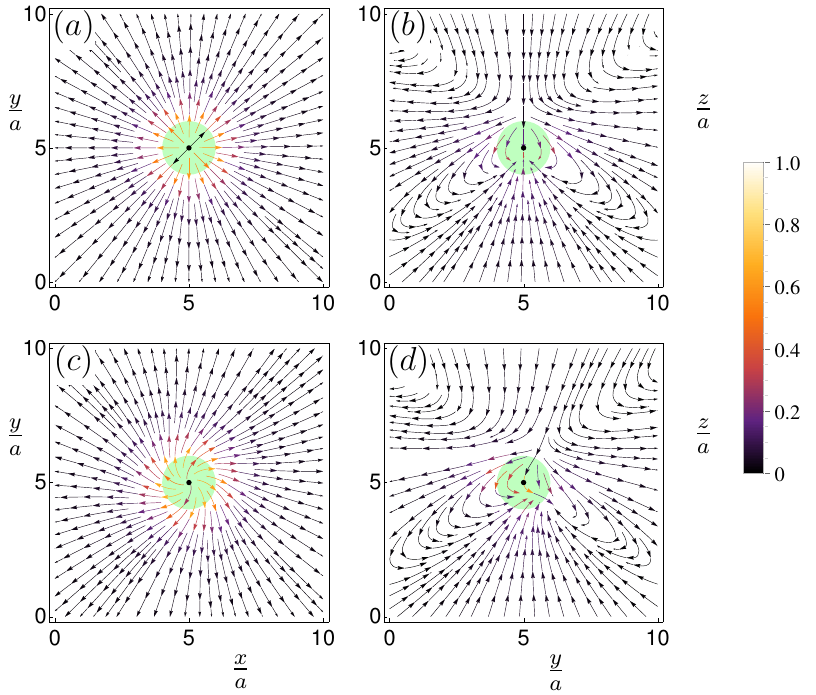}
  \caption{(Color online) 
  Velocity flow fields are generated in the lab frame by a squirmer (top panels) and a chiral squirmer (bottom panels). The velocity profiles are projected on $xy-$ plane for $z = a$ (left panels), and on $yz-$ plane for $x = a$ (right panels). The planes are shown to touch the squirmer (indicated in green) at the black dot. The parameter values are the same as in Fig.~\ref{fig:slip}. The color code provides the magnitude of the flow field.}
   \label{fig:flowfield}
\end{figure}

Fig.~\ref{fig:flowfield} depicts examples of the lab frame flow field generated by a simple squirmer (top panels (a), (b)) and by a chiral squirmer (bottom panels (c) (d)). The flows shown are projected on planes that touch the squirmer at one point (black dots). The flow field of a simple squirmer is axially symmetric concerning the $z$-axis, i.e., the direction of propulsion, (Fig.~\ref{fig:flowfield}(a)). The flow pattern in the $yz$-plane is mirror symmetric to the $x$- and $y$-axes (Fig.~\ref{fig:flowfield} (a),(b)). In contrast, the chiral squirmer produces a rotational flow around the $z$-axis with clockwise sense of rotation when viewed from the top  for $z > 0$ (Fig.~\ref{fig:flowfield}(c)) and counterclockwise for $z < 0$ (not shown). 
Note that for $z = 0$, no chiral component exists in the flow.
In the $yz$-plane, there is no mirror symmetry as a result of chirality 
(Fig.~\ref{fig:flowfield}(d)). Also note that the $l = 1$ modes do not generate chiral flow patterns. However, they generate a body rotation ${\bf \Omega}$ of the chiral squirmer. Chiral contributions to the flow stem from terms with $l\geq 2$ such as the contribution with the term $\sim \gamma_{20}^r/r^3$ in Eq.~(\ref{eq:VelocityField}). The dominant term in the far-field is proportional to $\sim \beta_{20}^r/r^2$, which corresponds to a Stokes doublet and implies the action of a force dipole on the fluid \cite{Blake}. Thus, for $\beta_{2 0}^r/\beta_{1 0}^r > 0$ the chiral squirmer is a puller and for  $\beta_{2 0}^r/\beta_{1 0}^r < 0$, it is a pusher. 
For an illustration, see the 3D flow field of a puller and pusher in Fig.~\ref{fig:chi-lambda}(a) and (b).

\subsection{Path of a chiral squirmer} 
\label{subsec:path}

\begin{figure}[t]
  \centering
  \includegraphics[scale=0.8]{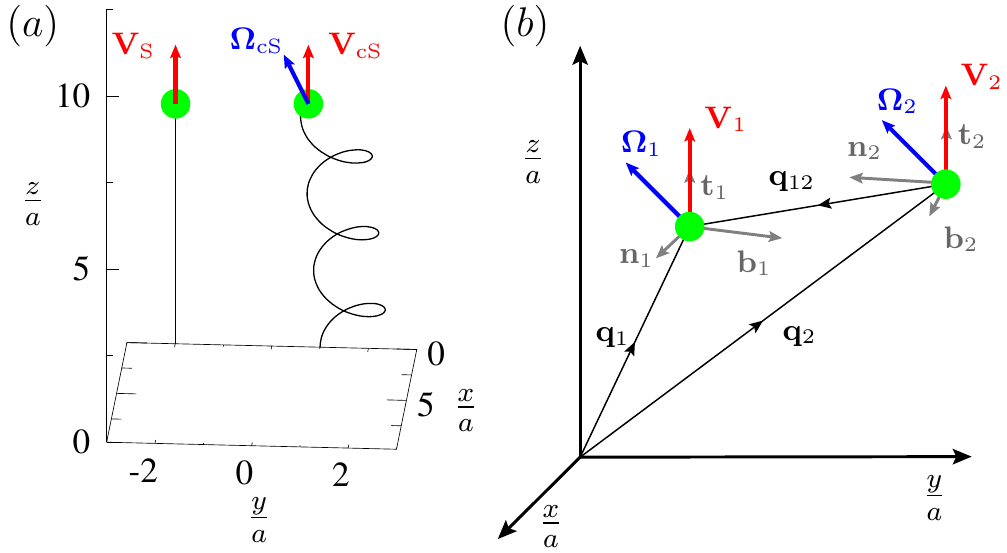}
  \caption{(Color online) 
    (a) Swimming path of a squirmer (straight line) and a chiral squirmer (helix) 
    for the same parameter values as in Fig.~\ref{fig:slip}. The swimming velocity
    and the rotation rate are indicated.
    (b) Schematic representation of two chiral squirmers in the laboratory frame of reference.}
  \label{fig:path_hydro}
\end{figure}

The equations of motion of the chiral squirmer determining the swimming path ${\bf {q}}(t)$ and its instantaneous orientation $({\bf n}$,   ${\bf b}$,  ${\bf t})$ read in the lab frame,
\begin{align}
\label{eq:CSq_motion}
{\bf \dot{q}} = {\bf V}\,,\,\, 
 \left[\begin{array}{c} {\bf \dot{n}} \\ {\bf \dot{b}} \\{\bf \dot{t}} \end{array}\right]  = {\bf \Omega} \times 
 \left[\begin{array}{c} {\bf n} \\ {\bf b} \\ {\bf t} \end{array}\right]\,,
\end{align}
where the dots denote time derivatives.
For time-independent coefficients $\beta_{lm}$ and $\gamma_{lm}$, ${\bf V}$ and ${\bf \Omega}$
are constant when described in the body frame.
The angle $\chi$ between ${\bf V}$ and ${\bf \Omega}$ obeys  
$\chi = \cos^{-1}\left(\frac{{\bf V}\,\cdot\,{\bf \Omega}}
{|{\bf V}||{\bf \Omega}|}\right)$.  
For ${\bf V} \parallel {\bf \Omega}$, we get $\chi = 0$, and 
the resulting swimming path is a 
straight line, see Fig~\ref{fig:path_hydro}(a). 
In this case, the swimmer rotates around the axis of motion. 
For $\chi = \pi/2$, the chiral squirmer moves in a circular path in a plane (not shown). 
For other values of $\chi$, the path of a chiral squirmer is a helix, see
Fig~\ref{fig:path_hydro}(a).
Using the velocity and rotation rate of the swimmer, one can calculate 
the curvature $\kappa_0 = |{\bf \Omega} \times {\bf V}|/|{\bf V}|^2$ and 
the torsion $\tau_0 = |{\bf \Omega} \cdot {\bf V}|/|{\bf V}|^2$ or
alternatively the radius $r_0 = \kappa_0/(\kappa^2_0 + \tau^2_0)$
and the pitch $p_0 = \tau_0/(\kappa^2_0 + \tau^2_0)$ of the swimming path.
Without loss of generality, we assume that the squirmer 
rotates in the $\mathbf{n - t}$ plane, i.e., we set $\gamma_{11}^i = 0$ in Eq.~(\ref{eq:vel_rot}), with the magnitude $| \bm \Omega| = v/a$. 
With this choice, we can write its components as 
$\gamma_{11}^r/a = (v/a)\sin \chi \,\, , \gamma_{11}^i/a = 0$ and $\gamma_{10}^r/a = (v/a)\cos \chi$.

\section{Hydrodynamic interaction of two chiral squirmers} 
\label{sec:2chsquirmer}

Having discussed the flow field and motion of a chiral squirmer in the lab frame, we can now focus on the motion of a pair of hydrodynamically interacting chiral squirmers. We consider the situation where the swimmers are far apart than their diameter; subsequently, the superposition of the flow fields can provide an excellent approximation to the combined flow and permits to study hydrodynamic interactions \cite{Pedley}. As a result, a given chiral squirmer obtains an additional contribution to its velocity and rotation rate due to the flow field induced by the other swimmer. 
Thus, the equations of motion for swimmer one in presence of swimmer two read
\begin{align}
  \label{eq:motionHydro}
 {\bf \dot{q}}_1
  & = {\bf V}_1 + {\bf u}_{2}({\bf q}_{1 2}, {\bf n}_2, {\bf b}_2, {\bf t}_2) \nonumber\\
  \left[\begin{array}{c}  {\bf \dot{n}}_1  \\ {\bf \dot{b}}_1 \\{\bf \dot{t}}_1 \end{array}\right]  
  & = 
  \left[ {\bf \Omega}_1 + \boldsymbol{\omega}_{2}\Big({\bf q}_{1 2}, {\bf n}_2, {\bf b}_2, {\bf t}_2 \Big) \right]
  \times 
  \left[\begin{array}{c} {\bf n}_1  \\ {\bf b}_1 \\ {\bf t}_1 \end{array}\right] 
   \,,
\end{align}
where ${\bf u}_{2}({\bf q}_{1 2}, {\bf n}_2, {\bf b}_2, {\bf t}_2)$ and 
$\boldsymbol{\omega}_{2}({\bf q}_{1 2}, {\bf n}_2, {\bf b}_2, {\bf t}_2)$  
are the velocity field and vorticity created by swimmer two at the position of swimmer one, 
as in Eqs. (\ref{eq:VelocityField}) and ~(\ref{eq:VorticityField}).
The time dependent radial separation distance between the swimmers is given by 
$R = |{\bf q}_{1 2}| = |{\bf q}_1-{\bf q}_2|$, see Fig.~\ref{fig:path_hydro}(b).
A corresponding equation holds for swimmer two. 
Therefore, we see that the unperturbed velocity and rotation rate of a swimmer gets modified due to the velocity field 
and vorticity, respectively, of the other swimmer.

\begin{figure*}[t]
\centering
\includegraphics[scale=0.66]{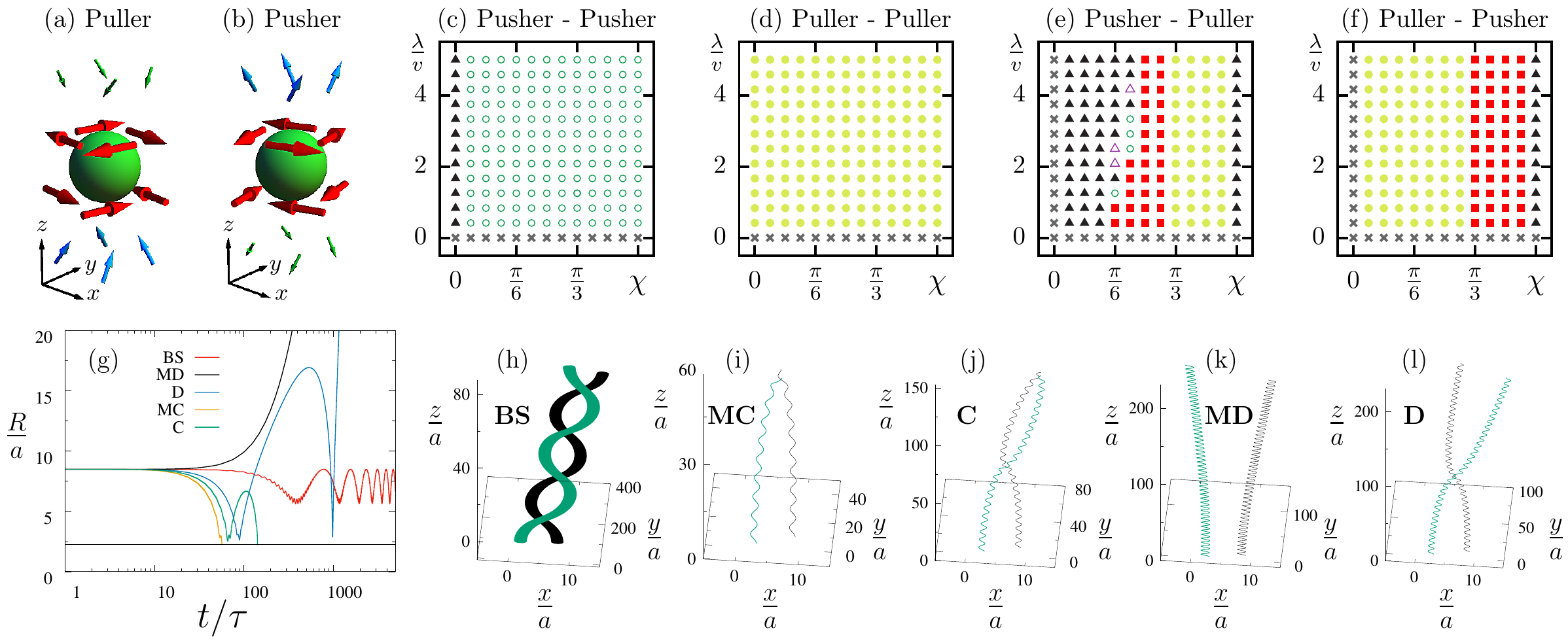}
\caption{(Color online)
(a)-(b) Flow around a puller and a pusher. 
(c)-(f) Represent the numerically obtained swimming behaviors (states) of a pair of hydrodynamically interacting chiral squirmers. 
Different symbols represent different states.
Square symbol - bounded state (BS), 
closed circles - monotonic divergence state (MD), 
open circles - divergence state (D), 
closed triangles - monotonic convergence state (MC), 
open triangles - convergence state (C), and 
cross symbols represent the situation where swimmers move parallel to each other.
The initial positions for the swimmers are set to 
${\bf q}_1 = (9,9,0) a$ and ${\bf q}_2 = (3,3,0) a$. 
Swimmers have the same 
initial velocity ${\bf V}_1 = {\bf V}_2 = v (0, 0, 1)$ 
and rotation rate ${\bf \Omega}_1 = {\bf \Omega}_2 = v\,(\cos \chi, 0, \sin \chi)/a$, 
which depends on the angle $\chi$.
(h)-(l) show the corresponding trajectories of the states, 
for the values $\chi = \pi/3$ and $(\lambda_1, \lambda_2) = v(1, -1)$ for BS, 
$\chi = 5 \pi/12$ and $(\lambda_1, \lambda_2) = v(-2, 2)$ for MC, 
$\chi =  \pi/6$ and $(\lambda_1, \lambda_2) = v(-2, 2)$ for C, 
$\chi = \pi/3$ and $(\lambda_1, \lambda_2) = v(-1, 1)$ for MD, 
and $\chi = 5 \pi/12$ and $(\lambda_1, \lambda_2) = v(-1, -1)$ for D.
(g) The corresponding radial separation distance $R$, 
scaled by the radius of the swimmer $a$, between the swimmers 
is plotted as a function of time $t$, scaled by $\tau = v/a$. }
\label{fig:chi-lambda}
\end{figure*}

Using Eq.~(\ref{eq:motionHydro}), we numerically calculate the trajectories of a pair of chiral swimmers and investigate their combined behavior, see Fig.~\ref{fig:chi-lambda}.
We consider chiral swimmers having translational velocities of equal magnitudes, i.e., 
$|\mathbf{V}_1| = |\mathbf{V}_2| = v$. The rotation rates of the swimmers are in general different and read, 
${\bf \Omega_1} = v(\sin\chi_1, 0,\cos\chi_1)/a$ for swimmer one and 
${\bf \Omega_2} = v(\sin\chi_2, 0,\cos\chi_2)/a$ for swimmer two. 
Modification in $\chi_1$ and $\chi_2$ changes the corresponding torsion and curvature of the swimmers' helical trajectories. 
Additionally, the flow field of one swimmer influences the motion of the other swimmer. As mentioned earlier, $l > 1$ modes in the velocity field, Eq.~(\ref{eq:VelocityField}), play a vital role in the hydrodynamic interaction between the swimmers. 
Thus, we choose the $l = 2$ modes corresponding to swimmer one as 
$3\beta_{20}^{r} = 3\gamma_{20}^{r} = \lambda_1$ and similarly for swimmer two as 
$3\beta_{20}^{r} = 3\gamma_{20}^{r} = \lambda_2$.
Note that for $\lambda_1 \neq \lambda_2$, the swimmers differ in their chiral flows that they generate.
Thus, variation in $\chi_i$ and $\pm \lambda_i \,(i = 1,2)$ determine the nature of the interaction between the chiral squirmers and gives rise to several interesting swimming characteristics.
We have considered various possible initial configurations for the swimmers. 
Here, we present only the planar configuration, 
where both the swimmers start initially on the $xy$-plane, 
separated by a distance $d$, moving in the positive $z$-direction. 
This particular choice of the configuration recovers the known behaviors exhibited by two simple squirmers (without chirality) 
and some additional exciting behaviors discussed below. 
Note that swimmers get enough time to interact in this configuration, whereas it may not be the case in other configurations.

\subsection{State diagram in the $\chi$ - $\lambda$ plane}
\label{subsec:chi-lambda}

We start from initial positions $\mathbf{q}_1 = (9,9,0)a$ and $\mathbf{q}_2 = (3,3,0)a$ of the squirmers 
with ${\bf t}_1$ and ${\bf t}_2$ in the $z-$ direction, and ${\bf n}_1$ and ${\bf n}_2$ in the $x-$ direction. 
Note that the separation distance between the swimmers is larger than the 
radius $a$ of the swimmers. 
We choose $\chi = \chi_1 = \chi_2$, i.e., the relative orientations of the swimmers with respect to their motion are the same, and $\lambda = |\lambda_1| = |\lambda_2|$, i.e., the strength of the hydrodynamic flow fields of both the swimmers are identical. 
Consider the sub cases, i.e., 
pusher - pusher: $(-\lambda_1, -\lambda_2)$, 
puller - puller: $ (\lambda_1, \lambda_2)$, 
pusher - puller: $(-\lambda_1, \lambda_2)$, and
puller - pusher: $ (\lambda_1, -\lambda_2)$.
Thus, by varying the strengths of $\chi$ and $\lambda$, we have observed different behaviors of two hydrodynamically interacting swimmers, see Fig.~\ref{fig:chi-lambda}. 
For example, (i) bounded state (BS), where the swimmers spiral around each other and move synchronously with a separation distance that changes periodically, (ii) monotonic convergence (MC), where the swimmers approach each other and reach a minimum distance at which near field interactions become more dominant over the far field interactions, (iii) convergence (C), where the swimmers attract each other after some transient behavior, (iv) monotonic divergence (MD), 
where the separation distance between the swimmers increases monotonically as they move together, and (v) divergence (D), where the swimmers diverge after a transient attractive behavior. 
Fig.~\ref{fig:chi-lambda}(c-f) depicts the detailed phase diagrams, 
the examples of the observed states are showed in the panels (h-l), 
and the corresponding radial separation distance as a function of time is plotted in the panel (g).

The peculiar bounded state is a result of the system's chiral nature, which can be observed only for the asymmetric combination of two swimmers, i.e., either puller - pusher or pusher - puller. Except for the bounded state, the other states have also been reported for the simple swimmers that move either in a straight line or a circular path \cite{Pedley,Lauga,Pagonabarraga}. These states can be observed either for the symmetric or asymmetric combinations of two swimmers. Note that when the swimmers exhibit the monotonic convergence or convergence behavior, the separation distance reduces as they approach each other. Below a certain distance, the far-field approximation becomes inaccurate, and the near-field becomes dominant. 
A rigorous analysis of the near-field interaction between two chiral squirmers, will be subjected to future work.

To investigate the robustness of the observed bounded states, we have considered various initial configurations of the swimmers. 
However, we have found that bounded states are observed only if the swimmers oriented approximately in parallel (side by side) configuration initially. For other initial conditions, the swimmers do not stay in proximity long enough 
to influence each other's motion.
Therefore, the contact time between the swimmers is shorter, and they do not exhibit 
the bounded state. 
Note that the observed bounded states are stable even with small perturbations to  $\lambda$ and $\chi$ 
which are $\epsilon_1 \sim 5 v/24$ and $\epsilon_2 \sim 0.01 \pi/24$, respectively. 
We have observed that the chiral paths but not the chiral flows influence the bounded state. 
Also, by setting $\lambda_{1,2} = 0$, i.e., in the absence of the $l = 2$ modes in the flow field, swimmers become neutral (neither pullers nor pushers), and they 
move in their respective directions without influencing each other's motion \cite{Pagonabarraga}.

\subsection{State diagram in the $\lambda$ - $\lambda$ plane}
\label{subsec:lambda-lambda}

\begin{figure}[hbt]
\centering
 \includegraphics[scale=1.1]{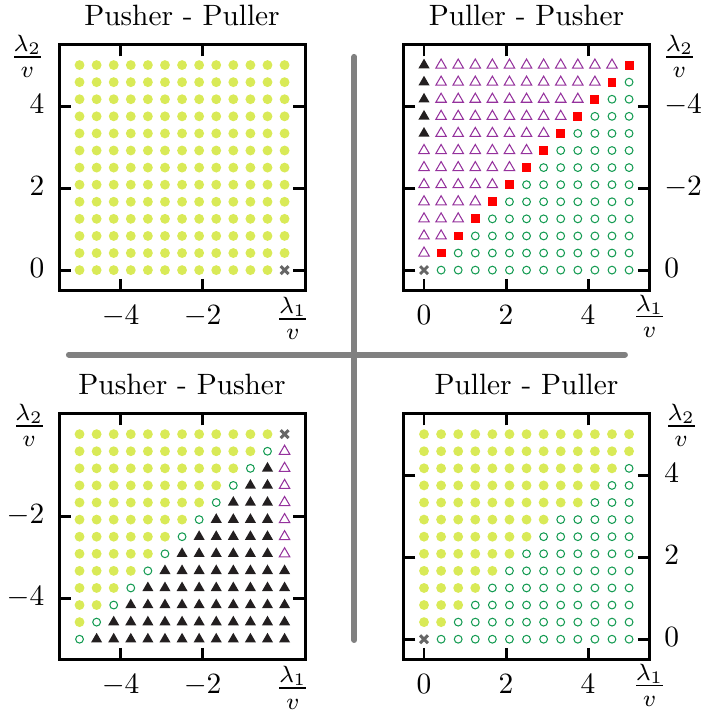}
\caption{(Color online) 
  Numerically obtained swimming behaviors of a pair of hydrodynamically interacting chiral squirmers, one with the strength $\lambda_1$ and the other one with $\lambda_2$. 
  Here, we set $\chi_1 = \chi_2 = \pi/3$. 
  Symbols are same as in Fig.~\ref{fig:chi-lambda}.}
\label{fig:lambda-lambda}
\end{figure}

In this case, we consider $\chi_1 = \chi_2 = \pi/3$ and study the effect of the hydrodynamic field  of one swimmer on the other. 
Fig.~\ref{fig:lambda-lambda} depicts the hydrodynamic behavior of two swimmers with varying $\lambda_1$ and $\lambda_2$. We found that interaction depends on the swimmer type, i.e., pusher or puller set by the sign of $\lambda_{1,2}$. 
For example, for the choice $\lambda = -\lambda_1 = \lambda_2$ swimmers exhibit a bounded state 
for the puller-pusher combination but show a monotonic divergence state for the combination of pusher-puller (see Fig.~\ref{fig:lambda-lambda}). 
This altered behavior is because of the asymmetry in the flow patterns about the direction of motion exhibited by chiral swimmers of puller and pusher nature, see Fig.~\ref{fig:chi-lambda}(a),(b).
Note that the flow pattern of an axisymmetric puller is mirror symmetric, about the swimming direction, to the flow pattern of its counterpart, a pusher. 
However, in chiral swimmers, the flow patterns of puller and pusher are not mirror symmetric to each other. 
Thus, by choosing an initial configuration, e.g., puller-pusher gives a different behavior 
compared to the puller-pusher combination. The same is reflected in 
Fig.~\ref{fig:lambda-lambda}.
Note that only the chiral swimmers with initial configurations, e.g., puller-pusher or pusher-puller, exhibit bounded states but not non-chiral swimmers. 
For the other combinations, i.e., pusher-pusher or puller-puller, depending on the strength of $\lambda_1$ and $\lambda_2$, a pair of chiral swimmers exhibit the states such as C, MC, D, and MD. Similar states have been reported in the case of a pair of axisymmetric swimmers \cite{Pedley, Lauga, Pagonabarraga}. 
Also, note that the observed bounded states are stable even with a small perturbation $\epsilon_1 \sim 5/24$ 
to $\lambda_1$ and $\lambda_2$.

\subsection{Phase diagram in the $\chi$ - $\chi$ plane}
\label{subsec:ss}

\begin{figure}[hbt]
\centering
\includegraphics[scale=1.1]{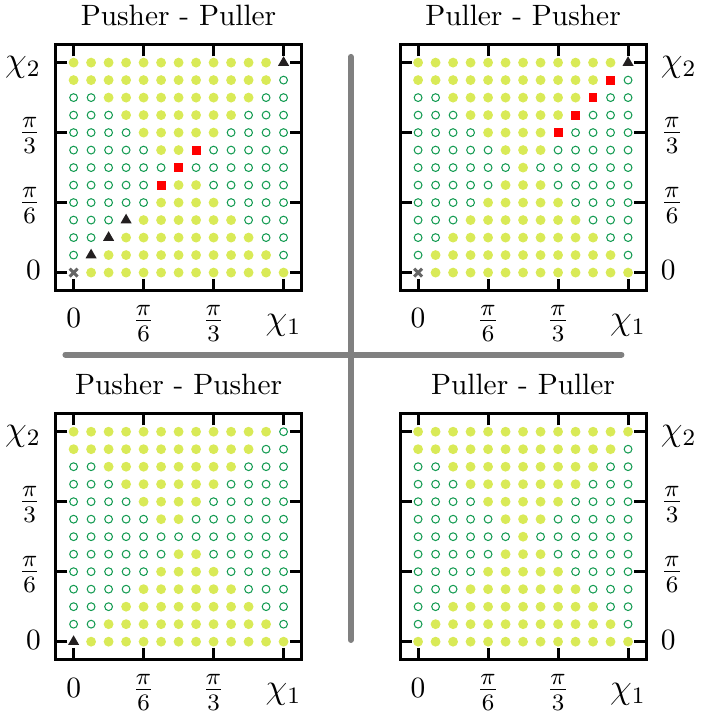}
\caption{(Color online)
  Numerically obtained swimming behaviors of a pair of hydrodynamically interacting chiral squirmers, one with the orientation $\chi_1$ and the other with $\chi_2$. 
  Here, we set $|\lambda_1| = |\lambda_2| = v$. 
  Symbols are same as in Fig.~\ref{fig:chi-lambda}.}
\label{fig:chi-chi}
\end{figure}

Here, we study the interaction between two swimmers by varying their respective initial angles between velocity and rotational axis, i.e., $\chi_{1,2}$ at time $t = 0$, by setting $\lambda_1 = \lambda_2 = v$. 
Varying $\chi$ induces different directions of the rotational axis with respect to the velocity of the swimmer. Note that $\chi$ does not influence the magnitude of swimmers' rotation rate; however, it influences the velocity field and vorticity. 
Therefore, when two swimmers interact hydrodynamically, their collective behavior is 
dictated by their respective $\chi$ values. This gives rise to different swimming states, see Fig.~\ref{fig:chi-chi}. 
In a puller-pusher or pusher-puller combination, if the net rotation rates of both the swimmers 
are equal, i.e., $\chi_1 = \chi_2$, the swimmers get attracted to each other, giving 
rise to either bounded or monotonic convergence states. 
If the net rotation rates are different, the swimmers mainly repel each other, giving rise to divergence and monotonic divergence states. Note that the observed bounded states are stable even with a small perturbation $\epsilon_2 \sim 0.01 \pi/24$ 
to $\chi_1$ and $\chi_2$. 
Interestingly, apart from the case $\chi_1 = \chi_2$, interaction of the squirmers flips about $\chi_1 = \chi_2 = \pi/4$ and shows symmetric pattern, see Fig.~\ref{fig:chi-chi}.

\section{Conclusions}
\label{sec:conclusions}

Using the chiral squirmer model, a generalization of the well-known squirmer model, we have investigated the dynamic behavior of two chiral squirmers coupled hydrodynamically. 
A chiral squirmer exhibits a chiral asymmetry of the surface slip velocity due to which it has both linear and rotational motion. The coupling of linear and rotational motion of the swimmer leads to a helical swimming path. We found that when a pair of chiral squirmers interact hydrodynamically, they can exhibit various types of motion. 
We first re-established the well-studied behaviors in the case of a pair of simple axisymmetric squirmers \cite{Pedley, Lauga, Pagonabarraga}. 
However, we found situations in which two swimmers entered a bounded state and moved jointly on helical paths. This situation is related to swimmers' chiral nature, and we found it only for swimmers with different chiral flow patterns. 
These behaviors might be related to the planar oscillatory movements reported for three-sphere swimmers, which also stem from rotational contributions in the hydrodynamic interactions but lack chirality \cite{Yemons}. 
Bound states of rotating swimmers have been previously reported for a pair of \textit{Volvox} algae that rotate and interact hydrodynamically near a substrate in the presence of gravity \cite{drescher}. In this case, the observed bounded states are due to the combined effect of hydrodynamic interaction between the spinning bodies, lubrication force between them when they close by, and the gravity.
However, in the current study, the observed bounded states are because of the chiral nature of the swimmers and their corresponding helical trajectories.
The significance of bounded motion lies in its possible role in fertilization in an adverse environment. Notably, throughout this work, we assume constant surface squirming motion or slip velocity. However, in general, 
microorganisms may alter their squirming velocity in the presence of a nearby microswimmer. 

Our model could be applicable to study the migration and the collective behavior of ciliated microorganisms and artificial swimmers \cite{ismagilov, golestanianart, dreyfus, paxton, hogg}. It can also be further extended to study Chemotaxis \cite{FriedrichPNAS2007,FriedrichPRL2009, maity} or Phototaxis \cite{Jekely, Drescher_phototaxis}, where the hydrodynamically coupled active swimmers move against a chemical gradient or a light source. As in the case of sperm cells \cite{FriedrichPNAS2007,FriedrichPRL2009}, the presence of an external stimulus may regulate the amplitudes of the slip velocity in the chiral squirmer model and may ultimately lead to steering towards a stimulus.

\acknowledgements

This work was supported by the Max Planck society and the Indian Institute of Technology Kharagpur, India.

\end{document}